\documentclass[twocolumn,trackchanges]{aastex62}
\usepackage{gensymb}
\usepackage{amsmath} 
\usepackage{gensymb}
\usepackage{wasysym}
\usepackage{xcolor}
\usepackage[inline]{enumitem}

\definecolor{cocoabrown}{rgb}{0.82, 0.41, 0.12}
\received{July 1, 2016}
\revised{September 27, 2016}
\accepted{\today}
\submitjournal{ApJ}

\shorttitle{An in-situ interplanetary ``U-burst": Observation and results}
\shortauthors{Mart\'inez Oliveros et al.}

\begin{document}

\title{An in-situ interplanetary ``U-burst": Observation and results}

\correspondingauthor{Juan Carlos Mart\'inez Oliveros}
\email{oliveros@ssl.berkeley.edu}

\author[0000-0002-2587-1342]{Juan Carlos Mart\'inez Oliveros}
\affil{Space Sciences Laboratory, University of California, Berkeley, CA 94720-7450, USA}

\author[0000-0002-5330-3131]{Saida Milena D\'iaz Castillo}
\affil{Observatorio Astron\'omico Nacional, Universidad Nacional de Colombia, Cra. 30 \# 45-03, 111321, Bogot\'a, Colombia}
\affil{Space Sciences Laboratory, University of California, Berkeley, CA 94720-7450, USA}

\author[0000-0001-6185-3945]{Vratislav Krupar}
\affil{Universities Space Research Association, Columbia, MD 21046, USA}
\affil{Heliospheric Physics Laboratory, Heliophysics Division, NASA Goddard Space Flight Center, Greenbelt, MD 20771, USA}
\affil{Department of Space Physics, Institute of Atmospheric Physics, The Czech Academy of Sciences, Prague 14131, Czech Republic}

\author[0000-0002-1573-7457]{Marc Pulupa}
\affil{Space Sciences Laboratory, University of California, Berkeley, CA 94720-7450, USA}

\author[0000-0002-1989-3596]{Stuart D. Bale}
\affil{Physics Department, University of California, Berkeley, CA 94720-7300, USA}
\affil{Space Sciences Laboratory, University of California, Berkeley, CA 94720-7450, USA}

\author[0000-0002-5041-1743]{Benjam\'in Calvo-Mozo}
\affil{Observatorio Astron\'omico Nacional, Universidad Nacional de Colombia, Cra. 30 \# 45-03, 111321, Bogot\'a, Colombia}


\begin{abstract}

We report and examine the observation of an unusual and rare in-situ electron observation associated with a solar type III radio burst on 24 December 1996. This radio event was accompanied by high energy electrons, measured by the Solid State Telescope (SST) on-board Wind spacecraft. The type III radio emission started at $\approx$~13:10~UTC and was associated to a C2.1 GOES-class flare whose maximum was at 13:11~UTC and hosted by the active region NOAA 8007/8004, located on the west limb at N05$^\circ$ W74$^\circ$/N06$^\circ$ W85$^\circ$. During this event, the observation of an electron energy distribution likely to be associated with the radio emission was registered. The electrons arrive at the spacecraft predominantly from the anti-solar direction, suggesting that their general motion is Sunward along a closed magnetic field line. \citet{1999GeoRL..26.1089L} propose a model in which energetic electrons are injected into a coronal flux tube at one of its footpoints, releasing standard type-III emission.  As the magnetic field then directs them back toward the magnetic-conjugate footpoint of the first, the electrons release subsequent emission whose radio profile is a quasi-time reversal of the standard.  We have constructed a cylindrical flux-rope facsimile of this scenario that reproduces the U-burst profiles. We also report  observational features indicating a secondary electron energy distribution and propose a scenario that explains this feature.
\end{abstract}




\section{Introduction} \label{sec:intro}
Solar flares and Coronal Mass Ejections (CMEs) are examples of explosive transient phenomena \citep[e.g.][]{2008LRSP....5....1B,2011LRSP....8....1C,2012LRSP....9....3W}. Both phenomena show ``unique" characteristics in the decametric and hectometric dynamic radio spectra \citep[e.g.][]{2005JGRA..11012S07G,2014RAA....14..773R, 2017A&A...597A..77R,2017JSWSC...7A..37M}. Type II radio burst are slow drifting features in dynamic radio spectra commonly associated with CMEs \citep{1954Natur.173..532W}. It is believed that these are produced by CME-driven shocks which accelerate electrons inside the magnetic structure of the CME. These accelerated electrons subsequently drive Langmuir waves near the plasma frequency and through a non-linear process generating radio waves at the plasma frequency and its harmonics \citep{1954Natur.173..532W,1985srph.book..333N}. Solar flares are usually associated with frequency-drifting features commonly known as Type-III radio bursts.  It is accepted that type III radio emission is driven by beams of electrons accelerated by reconnecting magnetic flux in the solar corona.  These high-energy electrons follow the streamlines of the magnetic fields into which they are injected \citep{1954Natur.173..532W}. The radio emission thus excited is therefore heavily influenced by magnetic connectivity. U- and J-bursts are the observational evidence of such connectivity, which can be explained by electron beams propagating along closed magnetic field lines or flux loops \citep{1958Natur.181...36M,2017A&A...597A..77R}. How this radio emission is produced, and its relation to the local plasma frequency in the interplanetary medium, have been a topic of active research in recent decades \citep[e.g.][]{1958SvA.....2..653G,1998SoPh..181..363R,1998SoPh..181..395R,1998SoPh..181..429R,2002ApJ...575.1094W,2017A&A...597A..77R}.

The vast majority of U- and J-bursts are observed at high frequencies, signifying electrons that are relatively low in the corona \citep{2010RRPRA..15....5D}. However, there are instances of radio emission extending to much lower frequencies, and hence from apparently greater heights \citep{1999GeoRL..26.1089L,1995SSRv...71..231B}. Observations by \citet{1999GeoRL..26.1089L} of two U-bursts two hours apart on 1998 June 22 show radio emission, whose frequencies plummet to 1--1.5~MHz, before rebounding in the first instance to $\approx$2~MHz.  This suggests turn-around heights of 6 to 8 R$_{\astrosun}$, which roughly match the successive heights of a slow CME over the 2-hr interim.  They propose that the radio emission observed emanates from within the CME.  Energetic electrons associated with type III radio bursts have been detected  in-situ at 1 AU  by e.g. \citet{1981ApJ...251..364L}. They recognize two distinct classes in the electron energy populations they find, one associated with the impulsive phase of the flare, and the other following the first about a half hour thence (see also \citealt{1998ApJ...503..435E} and \citealt{1999ApJ...519..864K}). These interesting results lead us to ask what we can find out by applying the foregoing comparisons to U- and J-bursts.

In this article we compare radio observations and associated energetic electrons detected in situ by the Wind spacecraft on 24 December 1996. We use a simple model to explain our observations and suggest a possible physical scenario.

\section{Observations} \label{sec:observations}

\subsection{Radio emission} \label{observations:radio}

The Wind spacecraft is located at the first Lagrange point (L1) and contains several instruments to study the solar wind. In particular, the WAVES instrument on-board Wind was designed to study solar radio emissions and the characteristics of the interplanetary plasma \citep{1995SSRv...71..231B}. The Wind/WAVES instrument is composed of several receivers.  Two of these are employed to study solar radio emissions, the radio receivers RAD1 (frequency range 20--1040~kHz) and RAD2 (1.075--13.825~MHz).

\begin{figure}[htbp]
\centering
\includegraphics[width=\columnwidth]{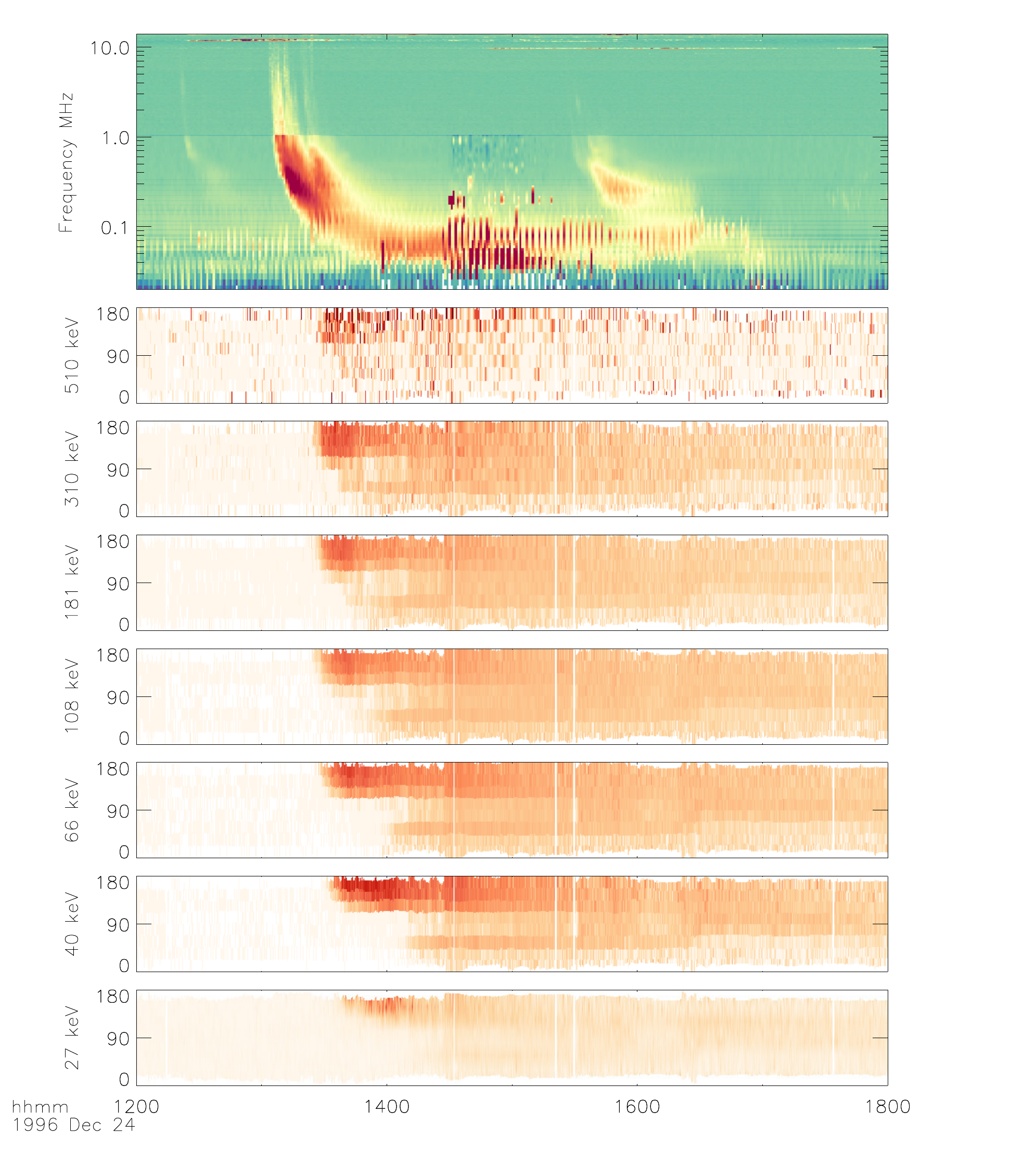}
\caption{Top: Wind/WAVES dynamic radio spectra. The six following panels show with the concurrent Wind/SST observations of the electron distributions for each energy channel as a function of time and pitch angle.}
\label{fig_radio}
\end{figure}

On 24 December 1996 RAD2, and subsequently RAD1, registered a type III radio burst (Figure~\ref{fig_radio}, top). This particular radio burst was accompanied by high energy electrons, measured by the Solid State Telescope (SST), also on board Wind \citep{1995SSRv...71..125L}. The type III radio emission started at $\approx$13:10~UTC and was associated with a C2.1 GOES class flare that reached maximum at 13:11~UTC and hosted by the active region NOAA 8007/8004, located on the West limb at N05$^\circ$ W74$^\circ$/N06$^\circ$ W85$^\circ$ \citep[see][for a full description of the radio event]{2008SoPh..253..305M}. The period between December 19-25 was characterized by a flurry of eruptive phenomena \citep{1999JGR...104.6679F}. In particular, on 19 December 1996 a CME event was detected at 16:30 UTC. According to Version 2 of LASCO CME List \footnote{\url{https://lasco6.nascom.nasa.gov/pub/lasco/status/Version2_CME_Lists/1996.12_CME_List.txt}} the CME propagated with an average velocity of about 332~km\,s$^{-1}$, a principal angle of 309 degrees and an angular width of 293 degrees. The CME propagated primarily in the SW direction and was composed by several concentric halos.

Wind/WAVES was developed with direction-finding capabilities, which means it is possible to determine the approximate location of the radio source generating the radio emission on the plane of the sky. To obtain the location of the type III radio burst observed on 24 December 1996, a modulation technique was applied to the data to retrieve the radio waves' direction of arrival \citep{1972Sci...178..743F} in the RAD1 range. Using this technique we were able to determine that the radio emission at frequencies above $\approx$200~kHz originated between $-$10$^\circ$ and +4$^\circ$ in longitude, where 0$^\circ$ refers to the line connecting Wind to Sun center.  At lower frequencies the direction of arrival was between $-$22$^\circ$ and $-$10$^\circ$ in longitude. This result places the sources over the West limb of the Sun as seen from Earth, therefore confirming the association of the radio source with NOAA active region 8007/8004 and the C2.1 GOES class flare.

\subsection{In-situ measurements} \label{observations:insitu}

Figure~\ref{fig_radio} shows the composite dynamic radio spectrum at the top, with the concurrent SST observations beneath. It is clearly seen in the SST data that high energy particles arrived at $\approx$13:30~UTC with pitch angles between 160 and 180 degrees. This indicates that the electrons arrived to the spacecraft from the anti solar direction. This electron distribution is observed in the majority of the SST energy channels. Later on, between $\approx$14:05~UTC and $\approx$14:30~UTC a second distribution of electrons was detected by SST, but in this case the pitch angle reached a maximum at about 30 degrees (solar direction).

\begin{figure*}[htbp]
\centering
\includegraphics[width=\linewidth]{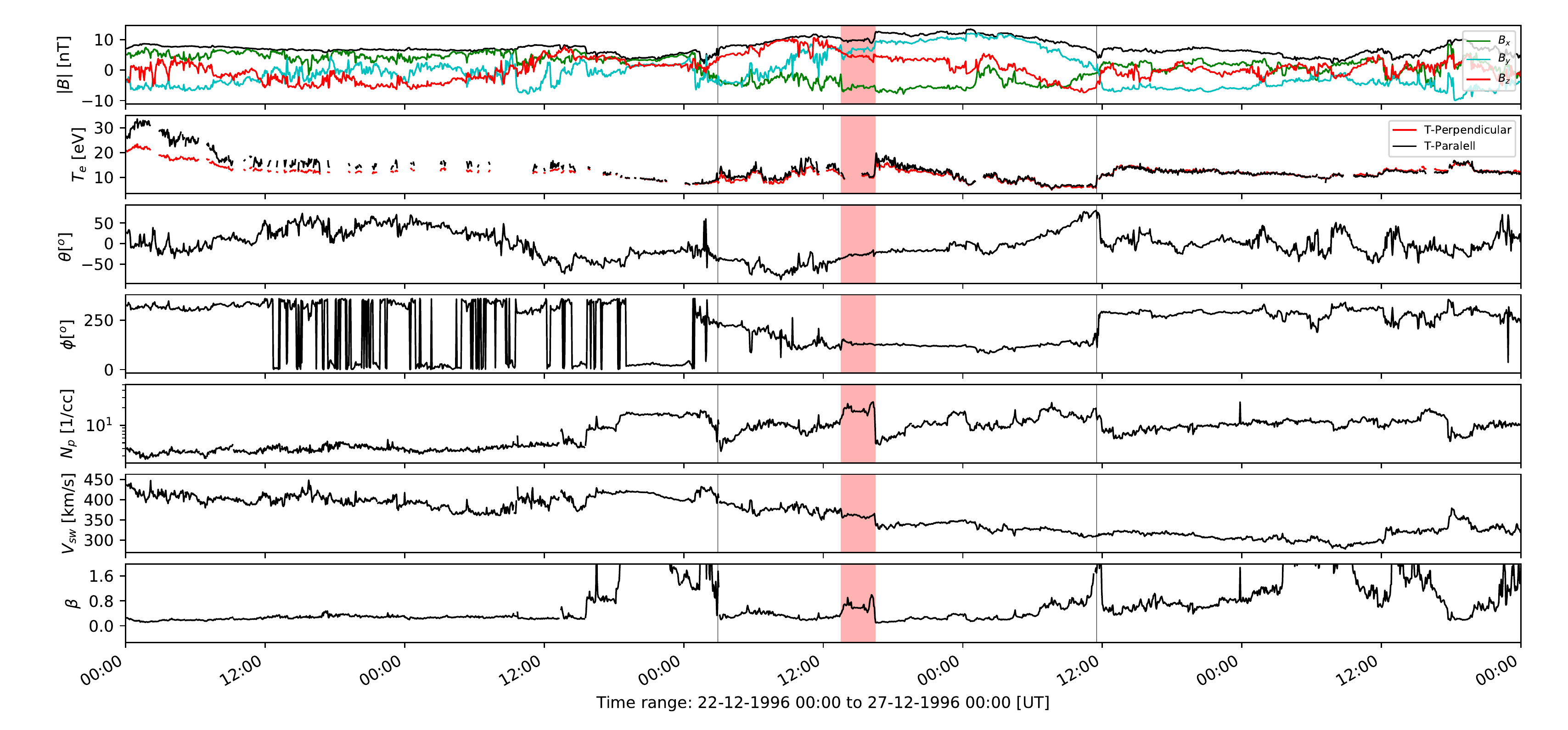}
\caption{In-situ data: Top: $B_x$, $B_y$, $B_z$, components and average magnetic field strength of the magnetic field, second panel from top: electron temperature, third panel from top: latitude of the magnetic field vector, fourth panel: longitude of the magnetic field vector (these two quantities show the magnetic field rotation), proton density, velocity and plasma $\beta$. Black lines represent the initial and final detection of the magnetic cloud studied and the red band corresponds to the electron beam in situ detection.}
\label{fig1}
\end{figure*}

Figure~\ref{fig1} shows principal characteristics of the plasma, measured in-situ (top to bottom: magnetic field strength, temperature, latitude and longitude of the magnetic field vector (magnetic field rotation), particle density, solar wind velocity and plasma $\beta$). We used the in-situ data (particles and magnetic
field)  to determine some model parameters. The key features of the in-situ observations can be summarized as follow:
\begin{enumerate*}[label={\alph*)}]
\item The initial magnetic cloud was detected on 1996-12-24T02:54 UTC;
\item the end of the magnetic cloud detection was approximately at 1996-12-25T11:30 UTC;
\item the maximum and  minimum values of the cloud magnetic field strength were 13.5 and 6 nT, respectively;
\item the mean solar wind speed: 356~km\,s$^{-1}$ and the mean proton density: 15 cm$^{−3}$;
\item the initial and final time of the electron beam in-situ detection was from 1996-12-24T13:30 UTC to 16:30 UTC ($\sim$3 hours).
\end{enumerate*}

\section{Proposed Scenario}\label{sec:scenario}

\subsection{In-situ features}

Following \citet{1999GeoRL..26.1089L}, we propose a scenario in which electrons accelerated by magnetic reconnection during the C2.1 flare traveled from a flaring footpoint along a magnetic loop contained within the CME while it was physically connected to that footpoint. In the following sections we explore this strong assumption of connectivity to explain the high energy electron observations. Based on the in-situ data (see section \ref{observations:insitu})  and the basic characteristics of the CME derived from the LASCO catalog (as described in section \ref{sec:intro}), we obtain the  observational constraints for our models. 

\begin{figure}[htbp]
\centering
\includegraphics[width=\columnwidth]{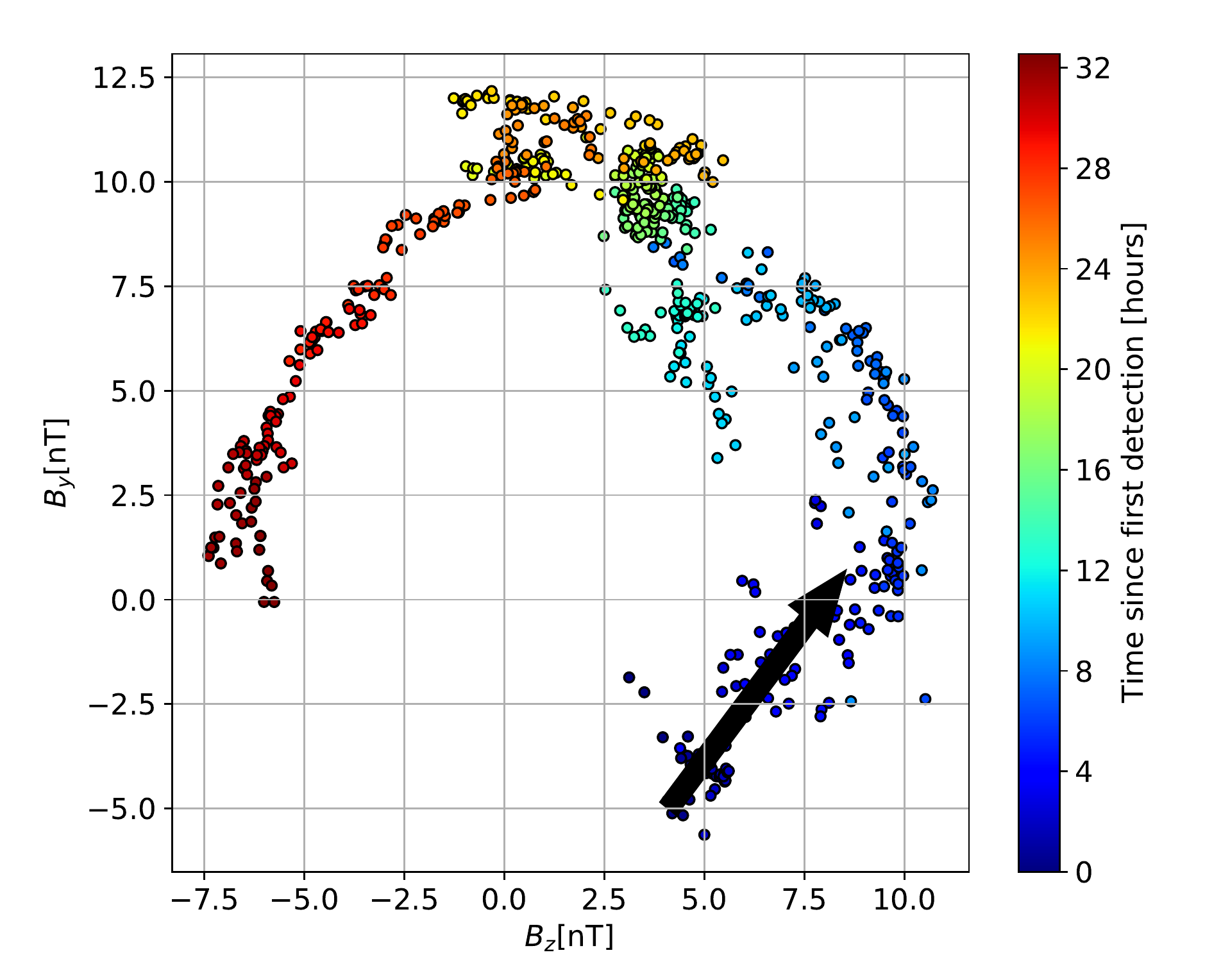}
\caption{In-situ $B_y$ and $B_z$ magnetic field projected onto the solar equator and plotted in GSE coordinates, as time progresses from 1996-12-24~02:54 (blue datapoints) to 1996-12-25~11:30~UTC (magenta) at 3-min intervals.  The magnetic-field projection is seen to rotate $\approx$225$^\circ$ from the GSE vantage.}
\label{fig2}
\end{figure}

We apply minimum variance analysis \citep[MVA; e.g.,][]{1967JGR....72..171S,1998AnGeo..16....1B} to the in-situ data to estimate the orientation of the flux rope axis at 1~AU (latitude and longitude in angular coordinates) and to verify the coherent rotation of magnetic-field vectors. Based on their convention, the flux rope helicity and magnetic cloud were classified as a North\,--\,East\,--\,South (NES) Right-handed from North ($+B_z$) to South ($-B_z$) and eastward $(+B_y$). Figure~\ref{fig2} plots the magnetic field projected onto the solar equatorial plane, in GSE coordinates $(B_z, B_y)$, as time progresses from 1996-12-24~02:54 (blue datapoints) to 1996-12-25~11:30~UTC (red-magenta).  The magnetic projection is seen to rotate 225$^\circ$ clockwise over this period and therefore confirming the observation of a magnetic cloud.

Assuming that the spacecraft passes through the magnetic structure close to its geometrical center, that the magnetic cloud has a circular cross section, and that there is not rapid expansion (quasi-stationary), it is possible to estimate the average size of the flux rope using Wind in-situ data.  Based on these assumptions, taking the average solar wind velocity to be 356~km\,s$^{-1}$ and ``passing" time of the magnetic cloud, we estimate the diameter of the magnetic flux rope, as $|\mathbf{v}_{sw}|\times\Delta t$ \citep{1998AnGeo..16....1B}. This calculation give us a lower limit for the magnetic cloud diameter of $\approx$0.27~AU.

\begin{figure}[htbp]
\centering
\includegraphics[width=\columnwidth]{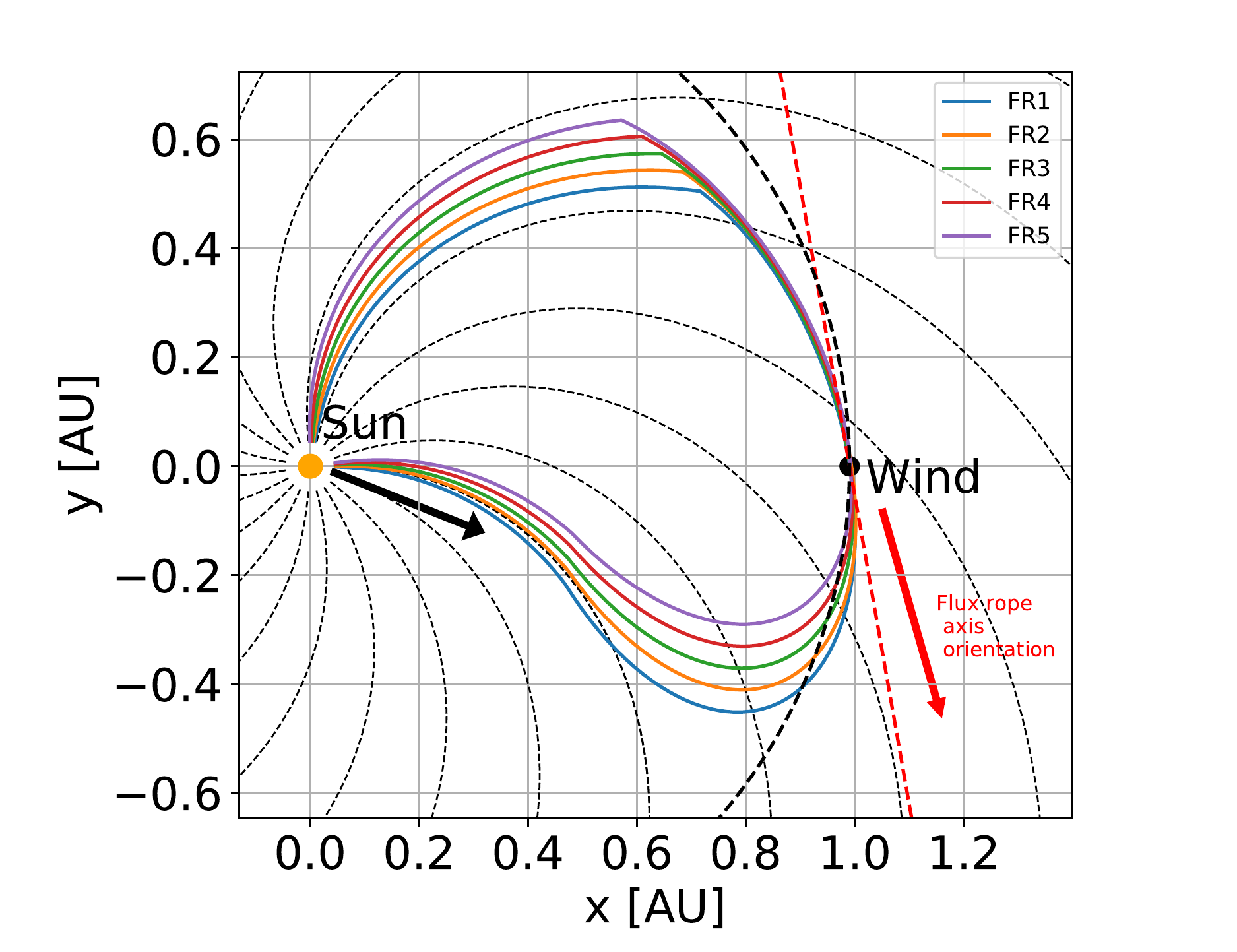}
\caption{Based on the MVA results, we propose a family of axis geometries for our flux rope model that fulfill the flux-tube axis orientation observed by Wind (dashed red line and red arrow). Outer extremities of colored magnetic loops conform to ellipses whose centers are marked by respectively colored dots.  In the first stages of the event, we considered the sense of electron beam movement in black arrow direction on the plot. The dashed thick black line represents Earth's orbit.}
\label{fig3}
\end{figure}

\subsection{Basic geometrical considerations}
Figure~\ref{fig3} show the suggested basic morphology of  the flux rope axis. It is composed by an elliptical front with its arms following the Parker spiral. The selection of the front width was made based on observations of other CMEs \citep{2008SoPh..253..305M}. The family of curves shown in Figure~\ref{fig3} represent different possible flux rope axes within the CME, based on the 19 December 1996 CME apparent morphology and assuming a constant expansion projected velocity of $\approx$332~km\,s$^{-1}$. 

The ellipses to which the fronts conform are specified by the position of the center of the ellipse, $(R_{HH}, \phi)$---where $R_{HH}$ expresses a heliocentric distance and $\phi$ an azimuth in the equatorial plane---the lengths, $a_{max}$, $a_{min}$, of its major and minor axes, and the the angle, $\Omega$, of the major axis with respect to the line of sight from the Sun to $(R_{HH}, \phi)$. We examine flux rope (FR) models whose inclinations with respect to the central line Sun-Wind have the values in the ordered set,  (5$^\circ$,8$^\circ$,11$^\circ$,14$^\circ$,17$^\circ$), to which we attach the model designations (``FR1,'' ``FR2,'' ``FR3,'' ``FR4,'' ``FR5''), respectively (see Figures 4 and 5).These inclinations are defined as steps of 3$^\circ$ around the average inclination of the CME, 11$^\circ$. The individual magnetic-flux lines in each of these modes are twisted, hence a set of helical curves which spiral about the axis of each model.

Table~\ref{tab:table1} summarizes the results and key parameters of the modeled flux rope magnetic field lines. The model was run with five different parameters (FR1--FR5), giving different total axial lengths. To better understand the paths followed by the particles, we recognize two separate components of the flux rope:  1.  Sun-Wind (SW), consisting  of  the  western  arm  of  the  flux  rope,  and  2. Wind-Sun (WS), corresponding to eastern arm, returning to the eastern footpoint thereof.  In this scenario, the electron population with low-pitch angle electrons proceeding outward from the Sun arrive at Wind at $\approx$14:00. These electrons are a ``kinetic echo" of the ingoing electrons that passed the Wind spacecraft location at $\approx$13:30  UTC. This electron population propagated inward  having  undergone  a reflection within the flux tube at some distribution of locations where the magnetic mirror is located.  

\subsection{Flux rope modeling}

The magnetic flux attached to the flux ropes in models FR1--FR5 is azimuthally symmetric and force free with force-free constant $\alpha$, confined within tubes of circular cross section whose outer radii are a function of heliospheric distance, $R$
\citep{1988JGR....93.7217B,1990JGR....9511957L,2009AnGeo..27.4057O}.

The models are fixed by a set of parameters that determine the profiles of $R$, the magnetic flux density, $B_z$, and its twist, $\alpha$ as a function of length along it.  Following \citet{2009AnGeo..27.4057O} we define a parameter $Y$ to be the ratio of the radial distance, $\rho$, of a point in the flux rope from the magnetic axis to the full radius, $\rho_0$, of the flux rope at distance $R$.

\begin{figure*}[htbp]
\centering
\includegraphics[width=0.45\textwidth]{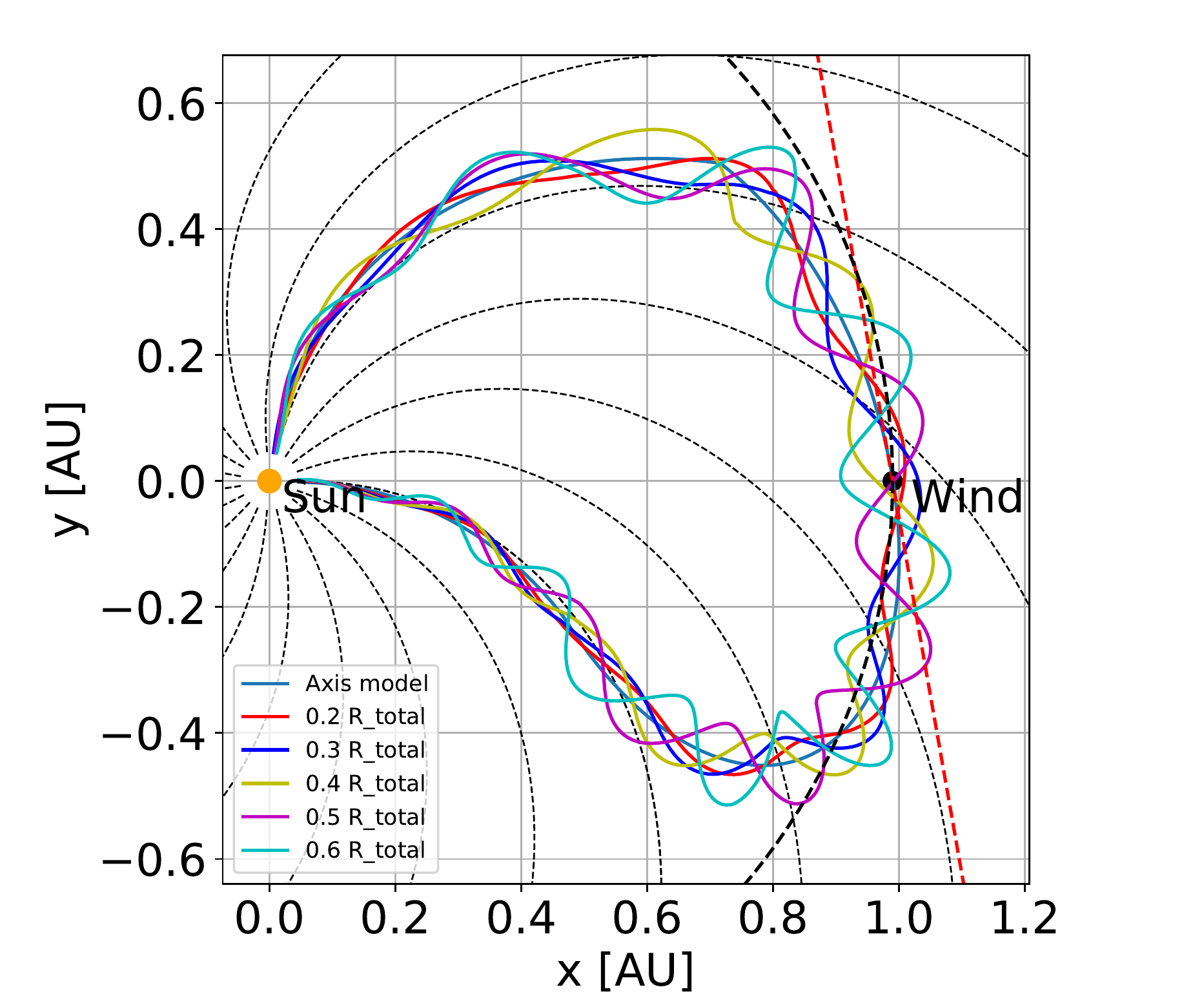}
\includegraphics[width=0.45\textwidth]{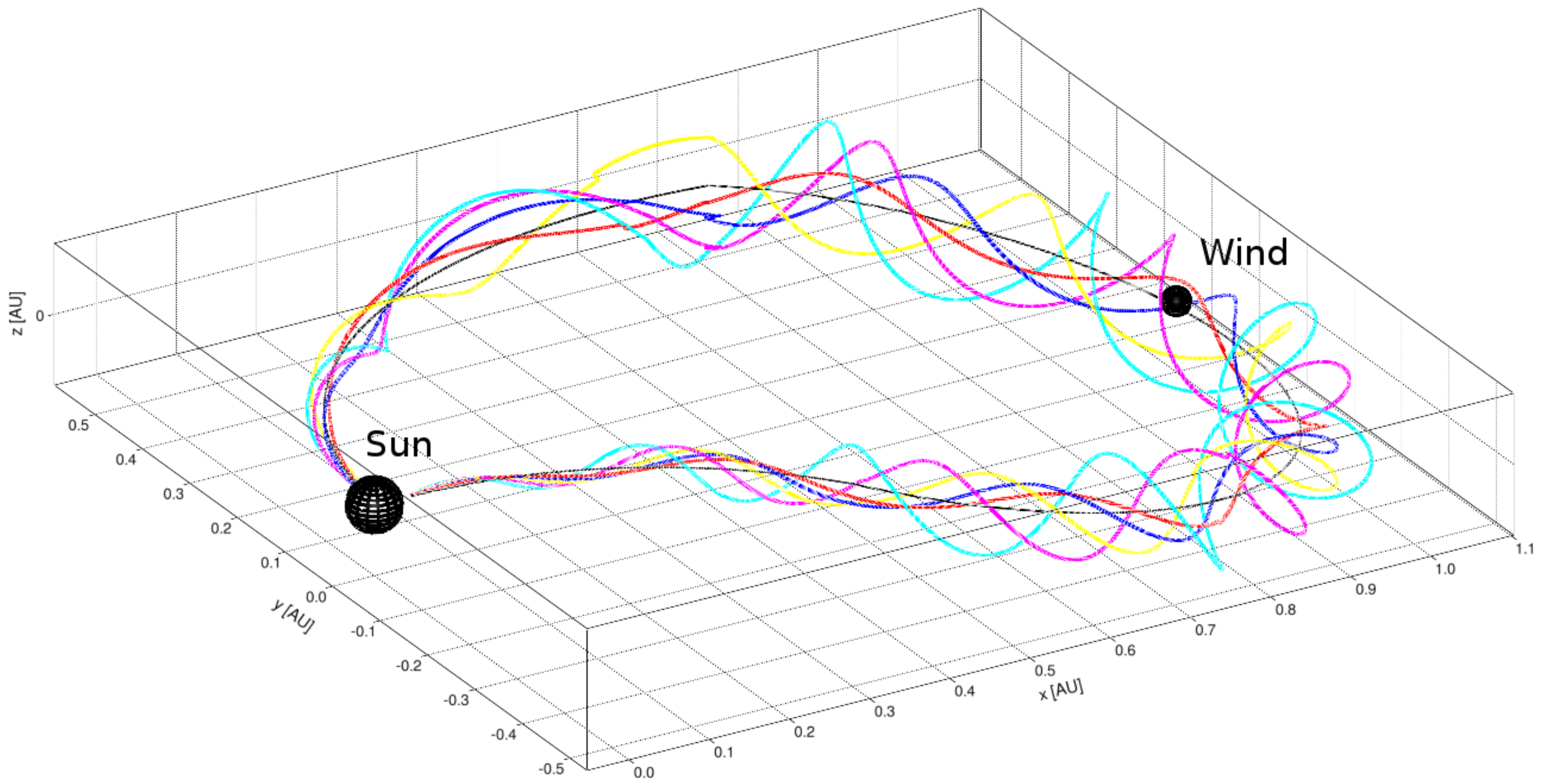}
\caption{Left: Magnetic streamlines of the FR1 model projected perpendicularly onto the ecliptic plane (left) and obliquely with respect thereto (right). The black curve traces the axis of the flux tube.  The colored loci trace individual magnetic streamlines attached to respective values of $Y$, listed from 0.2 to 0.6 in the legend at the lower left of the left plot.}
\label{fig4}
\end{figure*}

The primary condition to be satisfied by our models is the relative timing of arrivals of the various components of energetic electrons, from the onset arrival times of the in-situ electron influxes at the spacecraft within the magnetic cloud.  We assume that the magnetic cloud does not expand appreciably during its passage over the Wind spacecraft, and that the flux rope axis is the location over any perpendicular cross-section thereof at which the magnetic flux density is maximum. The electron arrival time gives us an idea of its location within the flux rope. This position in the flux rope axis is $Y$=0.4$\pm$0.2, telling us that the electrons were detected close to the center of the flux rope. Using the above, we limited our analysis to studying field lines with values of $Y$ between 0.2 and 0.6.

\begin{table*}[tbp]
\centering
\begin{tabular}{|c|c c c |c c |c c |c c |c c |c c|}
\hline 
\multicolumn{1}{|c}{Model}&
\multicolumn{3}{|c}{Length} &
\multicolumn{2}{|c}{$Y$=0.2} &
\multicolumn{2}{|c}{$Y$=0.3} &
\multicolumn{2}{|c}{$Y$=0.4} &
\multicolumn{2}{|c}{$Y$=0.5} &
\multicolumn{2}{|c|}{$Y$=0.6}\\
 & Total&
$\mathrm L_{SW}$ &
$\mathrm L_{WS}$ &
$\mathrm L_{SW}$ &
$\mathrm L_{WS}$ &
$\mathrm L_{SW}$ &
$\mathrm L_{WS}$ &
$\mathrm L_{SW}$ &
$\mathrm L_{WS}$ &
$\mathrm L_{SW}$ &
$\mathrm L_{WS}$ &
$\mathrm L_{SW}$ &
$\mathrm L_{WS}$ \\
\hline
FR1 & 3.08 & 1.51 & 1.57 & 1.59 & 1.63 & 1.72 & {1.68} & 1.92 & {1.80} & 2.19 & {1.96} & 2.64 & {2.21} \\
FR2 & 3.07 & 1.45 & 1.62 & {1.54} & 1.68 & 1.65 & {1.74} &
1.84 & {1.87} & 2.12 & {2.02} & {2.52} & {2.31}\\
FR3 & 3.04 & 1.39 & 1.65 & {1.48} & 1.72 & 1.59 & {1.78} & 1.76 & {1.93} & 2.04 & {2.08} & {2.39} & {2.41} \\
FR4 & 3.03 & 1.33 & 1.70 & 1.42 & {1.77} & {1.52} & {1.83} & {1.68} & 1.99 & 1.95 & {2.15} & {2.29} & {2.50} \\
FR5 & 3.02 & 1.28 & 1.75 & {1.35} & 1.81 & {1.46} & {1.88} & {1.61} & {2.05} & {1.85} & {2.24} & {2.19} & {2.59}\\
\hline
\end{tabular} 
\caption{Characteristic axial lengths measured in AU for the 5 models. The total length (second column) was calculated on the axis of the flux rope. The particle path from the Sun to Wind (west arm, $L_{SW}$) and for the path of Wind to the Sun (east arm, $L_{WS}$), was calculated also on axis. Similar computations were made at different positions Y (0.2, 0.3, 0.4, 0.5, 0.6) in  the flux  rope axis. The models FR1 and FR2 were selected as the best candidates in the real scenario (position, distance and time comparison with the in-situ data).}
\label{tab:table1}
\end{table*}

Figure~\ref{fig4} plots selections of individual magnetic streamlines in the FR1 model for $Y$ values ranging from 0.2 to 0.6.  We favor the FR1 model since it provides the closest electron arrival time to the observations. The ratio of the flux rope radius to  heliocentric distance is $R_{\rm FR}/R = v_{exp}/v_{\rm sw}$, where $v_{\rm sw}$ is the local flux-tube velocity, i.e., the solar-wind velocity, and $v_{exp}$ is the rate of expansion of the flux tube \citep{2009AnGeo..27.4057O}. For a flux rope with radius 0.135~AU at 1~AU, the ratio  v$_{exp}$/v$_{sw}$=0.135. The measured average solar wind speed in December 24\,--\,28 was $\approx$356~km\,s$^{-1}$, implying an expansion velocity of the magnetic structure of $\approx$48~km\,s$^{-1}$.

Using these constraints it is possible to model the magnetic structure (field lines for each of the values of $Y$) for each axis (family of curves) and estimate the approximate length of the magnetic field lines. 

\subsection{Magnetic field model and magnetic mirror points}
Coronal electrons in a magnetic flux tube spiraling with a significant pitch angle toward a footpoint of the loop are generally reflected back upward, a manifestation of the ``magnetic-mirror effect'' \citep{1973ppp..book.....K}, at the point at which the magnetic field reaches a particular critical value.  The exact depth at which this happens for individual electrons that have yet to penetrate to the mirror point depends upon their pitch angle where the magnetic field is less than at the mirror point.  For simplicity, we  assume a single depth for a given electron energy, notwithstanding that for a given pitch angle the mirror point is independent of energy.  This assumption allow us to suggest a scenario in which the low pitch-angle in-situ electrons detected by Wind (Figure~\ref{fig_radio}), arriving at $\approx$14:00~UTC, are the mirror-point echos of those that passed by the spacecraft earlier, at $\approx$ 13:30~UTC, spiraling {\it towards} the Sun.  We will call the time delay between these arrival times the round-trip ``flight time'' of the electrons from the spacecraft to the supposed mirror point and back.

To determine the magnetic field strength of the mirror and its heliocentric distance, we used the standard form of the interplanetary magnetic field in cylindrical coordinates,

\begin{align*}
\vec{B}_{FR} &= B_{R} \hat{R} - B_{\phi}\hat{\phi}\\
\end{align*}

with $B_{R}$ and $B_{\phi}$ corresponding to the following expressions:

\begin{align*}
B_{R} &= {B}(R_0,\theta,\phi_0)\left(\frac{R_0}{R} \right)^2\\
B_{\phi}&= {B}(R_0,\theta,\phi_0) \left(\frac{\Omega R_0^2 \sin{\theta}}{V_{SW} R} \right)
\end{align*}

where $\mathrm{R_0}$ is the heliospheric height at the solar wind source surface (0.045~AU or 10R$_{\odot}$) and $\mathrm{\vec{B}_R(R_0, \theta, \phi_0)}$ the magnetic field strength at the solar wind source surface \citep[e.g.][]{2019AnGeo..37..299L}. With this in mind, to obtain a magnetic field strength of 10nT at 1~AU (see Figure~\ref{fig1}, top panel, black curve), the source surface magnetic field strength must be $\approx$5000nT or $\approx$0.05~Gauss. From the solar wind parameters we used a solar wind velocity of $\approx$400~km\,s$^{-1}$. The equation above is limited to the ecliptic plane ($\theta$ = 90$^{\circ}$) with $\Omega$ been the angular rotational speed of the Sun.

\begin{table*}[!htbp]
\centering
\begin{tabular}{|c|c|c c|c c|c c|c c|c c |}
\hline
\multicolumn{1}{|c|}{Energy}&
\multicolumn{1}{|c|}{MP} &
\multicolumn{2}{|c|}{Y=0.2} &
\multicolumn{2}{|c|}{Y=0.3} &
\multicolumn{2}{|c|}{Y=0.4} &
\multicolumn{2}{|c|}{Y=0.5} &
\multicolumn{2}{|c|}{Y=0.6}\\
\hline
 &  & R$_{dis}$ & B$_{str}$ & R$_{dis}$ &
B$_{str}$ & R$_{dis}$ & B$_{str}$ & R$_{dis}$ &
B$_{str}$ & R$_{dis}$ & B$_{str}$\\
\hline
27keV & 0.75 & 0.79 & 23 & 0.82 & 21 & 0.88 & 19 & 0.96 & 17 & 0.90 & 18\\
40keV & 0.90 & 0.67 & 29 & 0.73 & 25 & 0.82 & 21 & 0.84 & 20 & 0.97 & 16 \\
66keV & 0.84 & 0.72 & 26 & 0.77 & 24 & 0.85 & 20 & 0.90 & 19 & 0.96 & 17\\
108keV & 0.68 & 0.85 & 20 & 0.85 & 20 & 0.86 & 19 & 0.97 & 16 & 0.86 & 24 \\
181keV & 0.49 & 0.92 & 18 & 0.96 & 17 & 0.93 & 18 & 0.88 & 19 & 0.93 & 18 \\
310keV & 0.47 & 0.92 & 18 & 0.96 & 17 & 0.94 & 17 & 0.87 & 19 & 0.95 & 17\\
\hline
\end{tabular}
\caption{Heliocentric distances and magnetic field strength values for FR1 for each value of $Y$, measured in astronomical units and nT, respectively. MP: Mirror point distance, R$_{dis}$: Heliocentric height in each mirror point and B$_{str}$:magnetic field strength.}
\label{tab:table2}
\end{table*}

\begin{figure*}[htbp]
\centering
\includegraphics[width=\linewidth]{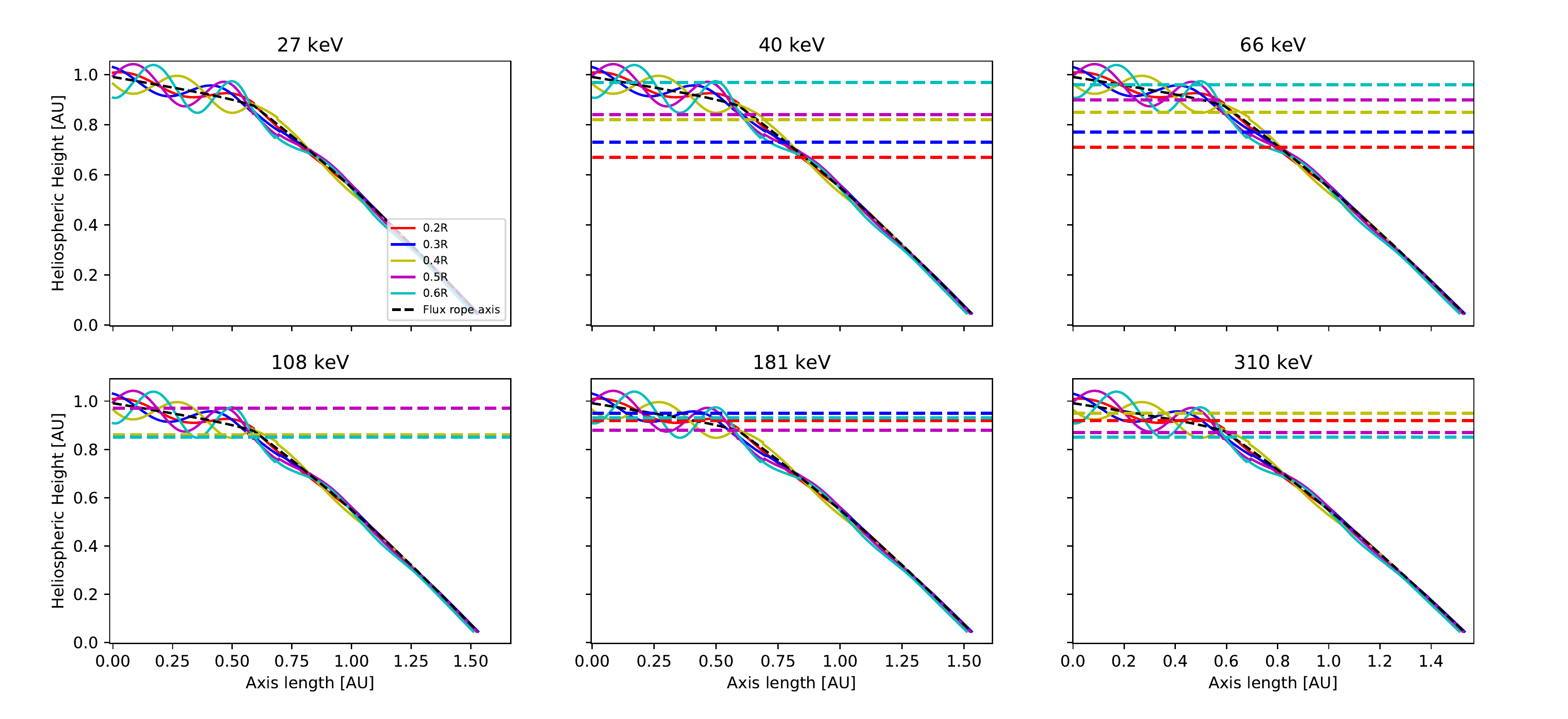}
\caption{Heliocentric height as a function of axis length for the FR1 as an example. Each plot shows the axis length and the magnetic field lines as a function of the heliocentric height. Each plot corresponds to different magnetic mirror point values (different energies). The dashed lines show the heliocentric height of the magnetic mirror point for different values of Y.}
\label{fig5}
\end{figure*}

Table~\ref{tab:table2} shows the calculated values of heliocentric distance and magnetic field strength at the magnetic mirror based on the adopted geometric flux rope  \citep{2009AnGeo..27.4057O} and interplanetary magnetic field model. Taking the flux-rope geometry of our favored model, FR1, it is relatively simple to derive the heliospheric location and magnetic field intensity at the mirror points for each electron population as a function of their flight times. Here we assume that the electrons propagate along the magnetic field line in a colissionless medium and without any kind of losses or diffusion. We acknowledge that these assumptions are not realistic, but demonstrate that within the uncertainty in the determination of the onset time of the lower pitch angle distribution, the electrons must have a mirror point located at the same location or very close. Figure~\ref{fig5} shows the heliocentric heights of the magnetic mirror points as horizontal lines laid on top of plots plots of the heights of magnetic streamlines as a function of axis length from the Wind. As there is a proportional relation between the distance to the mirror point and the energy of the electron distribution, the difference between the heliocentric distances can be explained by the different particles travel paths. The results demonstrate that our simple kinematic model can reproduce the time of flight of the electrons. Based on the modeling results and the observations, we suggest a simple scenario in which solar flare electrons released by the C2.1 GOES class flare on 24 December 1996 arrived at the spacecraft first and later propagated freely, without appreciable energetic losses in a near non-collisional environment. These electrons reached their respective mirror points depending of their energy and later on were detected again by the in-situ instruments onboard the spacecraft (Table~\ref{tab:table2}), and thus explaining the observations. 

\section{Discussion} \label{sec:discussion}

We follow the analysis of \citet{1999GeoRL..26.1089L} to develop a simple scenario based on a thin flux-rope model of CME magnetic clouds \citep{2009AnGeo..27.4057O} within which energetic electrons would spiral between magnetic-mirror points. This scenario was used to explain the in-situ observation of the 24 December 1996 event.

We find that the magnetic-mirror points for different energies cover a broad range of heliocentric distances, 0.67 to 0.9 AU, based on best-fit values of parameters that specify the geometrical architecture of the magnetic flux rope (see Figure~\ref{fig5}). However, we can safely say that the magnetic mirror is in general located at the same heliocentric distance at about 0.85 AU. These different distances are consistent with higher-energy electrons having shallower pitch angles, hence penetrating deeper sunward into the magnetic flux tube. This study, then, lends us a simple scenario of the magnetic geometry, including mirror-point locations, for each energy channel of the SST instrument with the mirroring magnetic field strength (see Table~\ref{tab:table2}).  

Finally, we question whether the observed event corresponds to a U-burst. The explanation of the radio U-burst nature is that this is the product of an electron beam propagating along closed magnetic field lines in the corona or interplanetary medium \citep{1958Natur.181...36M,2017A&A...597A..77R}. Because the mechanism of generation of the radio waves is plasma emission, the  frequency of the emission depends upon the density of the local medium. As the electrons spiral in the close magnetic loop, we see first a decrease in the emission frequency followed by a later increase.  In our case of the 24 December 1996 event, we see the type III radio burst in the dynamic spectra and simultaneously in-situ we observe the electron distribution likely to be associated with the radio emission. The fact that the electron distribution arrived to the spacecraft from the anti-solar direction allows us to say that we have ``seen" the electrons traveling in the direction of the Sun in a close magnetic field line.  This is the prescription for a U-burst. The subsequent observation of a second distribution of electrons can be explained by the reflection of the electrons at the magnetic mirror points.

\section*{acknowledgments}
We greatly appreciate the insights of Dr. Charles Lindsey. Work at UC Berkeley was supported in part by NASA grants NNX16AP95G and NNX16AG89G. V.K. acknowledges support by an appointment to the NASA postdoctoral program at the NASA Goddard Space Flight Center administered by Universities Space Research Association under contract with NASA and the Czech Science Foundation grant 17-06818Y. This paper uses data from the CACTus CME catalog, generated and maintained at the CDAW Data Center by NASA and The Catholic University of America in cooperation with the Naval Research Laboratory. SOHO is a project of international cooperation between ESA and NASA. 

\bibliographystyle{aasjournal}
\bibliography{uburst}

\end{document}